\documentclass[letter,nofootinbib,12pt]{revtex4-2}

\usepackage[colorlinks,citecolor=blue]{hyperref}
\usepackage{amsmath}
\usepackage{mathrsfs}
\usepackage{bbm}
\usepackage{amsfonts}
\usepackage{amssymb}
\usepackage{latexsym}
\usepackage{graphicx}
\usepackage[english]{babel}
\usepackage{multirow}
\usepackage{float}
\usepackage{url}
\usepackage{slashed}
\usepackage[compat=1.1.0]{tikz-feynman}
\usepackage{orcidlink}

\renewcommand{\Delta}{\varDelta} 
\renewcommand{\Gamma}{\varGamma} 
\renewcommand{\Omega}{\varOmega} 
\renewcommand{\Phi}{\varPhi} 
\renewcommand{\Psi}{\varPsi} 
\renewcommand{\Sigma}{\varSigma} 
\renewcommand{\Theta}{\varTheta} 
\renewcommand{\epsilon}{\varepsilon}

\newcommand{\be}{\begin{equation}}
\newcommand{\ee}{\end{equation}}
\newcommand{\ba}{\begin{array}}
\newcommand{\ea}{\end{array}}
\newcommand{\bea}{\begin{eqnarray}}
\newcommand{\eea}{\end{eqnarray}}

\begin{document}

\title{Amaterasu Cosmic Ray as Possible Support \\for
Electromagnetic Acceleration}

\author{Paul H. Frampton}
\email{paul.h.frampton@gmail.com}
\affiliation{Dipartimento di Matematica e Fisica 'Ennio De Giorgi',
Universit`a del Salento and INFN-Lecce,
Via Arnesano, 73100 Lecce, Italy} 

\bigskip
\bigskip

\vspace{2.0in}

\begin{abstract}
\noindent
The Amaterasu\footnote{named for a sun goddess in Japanese mythology.} cosmic ray particle announced in November 2023 is extraordinary. Its direction points back to the Local Void which contains no galaxies or known source. It is possible to show that the direction could not have been bent significantly by magnetic fields within the Milky Way.  This collection of facts constitutes a paradox.
In the present paper, we offer a resolution within the Electromagnetic Accelerating Universe (EAU) model in which a central r\^{o}le is played by charged Primordial Extremely Massive Black Holes (PEMBHs) whose Coulomb interactions underly accelerating expansion. Because structure formation of PEMBHs is electromagnetic while that for galaxies is gravitational, it is reasonable to expect PEMBHs inside the Local Void.
We provide an example where the cosmic ray primary is an antiproton and present it as supportiing evidence for the EAU model.

\end{abstract}


\maketitle 

\newpage

\noindent
{\it Introduction}.  Particle theory deals with very tiny particles which are typically smaller
than an atomic nucleus of size $10^{-15}$ m and therefore at least fifteen orders of magnitude
below the scales familiar to us. It treats objects far smaller than anything we can see with the
naked eye.

\bigskip

\noindent
Theoretical cosmology, by contrast, deals with very large objects which are typically larger than the Milky Way galaxy of size $10^{23}$ m and hence in excess of twenty-three orders of magnitude larger than familiar scales. It considers objects so huge that they stretch the powers of our human imagination.

\bigskip

\noindent
An outsider could reasonable surmise that physicists who research particle theory
form an entirely separate group than the physicists who research theoretical cosmology because the two groups study scales which differ by over thirty-eight orders of magnitude.
However, it has been known for many decades that this outsider's surmise is mistaken because when
we consider the early universe the temperature is so high that subnuclear particles are inevitable produced.
This fusion of the two research fields is sometimes displayed on an Ouroboros diagram, and the small-large connection has been very successfully exploited for over half a century.

\bigskip

\noindent
In the present article, we hope to convince the reader of the seemingly outrageous claim that the small-large connection exists {\it at the present time}. The claim is based on the recent observation of an extraordinary super-GKZ cosmic ray, called Amaterasu, which provides us
with a type of paradox whose resolution frequently results in a significant increase in human knowledge.

\bigskip

\noindent
Historically, he most important theoretical result for ultra high energy cosmic rays is the GKZ bound
\cite{Greisen:1966jv,Zatsepin:1966jv} that, to traverse the CMB, the energy is
bounded by
\begin{equation} 
E < 50 EeV
\label{GKZ}
\end{equation}
Observationally, over the years since \cite{Greisen:1966jv,Zatsepin:1966jv} the fortunes of the bound have ebbed and flowed as cosmic rays were claimed to 
violate Eq.(\ref{GKZ}) only for the data subsequently to be withdrawn. At present the cut off in Eq.(\ref{GKZ}) is very well established with only a few rare outliers exhibiting super-GKZ behaviour.

\bigskip

\noindent
The Amaterasu's energy is $E=240 EeV$, the third largest ever recorded after
previous super-GKZ cosmic rays with $320$ EeV (1991) and $E=280$ EeV (2001).

\bigskip

\noindent
What makes the Amaterasu particle\cite{abbasi} doubly
interesting is that not only is it super-GKZ, but the direction tracks back to
the Local Void which contains no galaxies and therefore,  it was thought wrongly, no source. To be sure that the direction in \cite{abbasi} cannot be misleading, the work of
Anchoroqui\cite{Anchordoqui:2018qom} was useful.

\bigskip

\noindent
The authors of \cite{abbasi} , however, restricted their attention to the
$\Lambda CDM$ model as the standard model of theoretical cosmology, without
considering the more recently proposed extension to the EAU-model
\cite{FramptonPLB,FramptonMPLA} in which the more than a century old assumption
\cite{Einstein} ,that gravitation dominates electromagnetism at all length scales greater
than molecules molecules, was called into question. As we shall analyse in the present paper,
the EAU-model provides a natural resolution of the Amaterasu paradox.

\bigskip

\noindent
{\it Electromagnetic Acceleration}. 
In the EAU-model, all the dark matter is composed of Primordial Black Holes (PBHs)
with that in galaxies and clusters being Primordial Intermediate Mass Black Holes (PIMBHs), while at galactic centres we have Primordial Supermassive Black Holes (PSMBHs).
All of these PBHs are electrically neutral like the stars and planets. Only
Primordial Extremely Massive Black Holes (PEMBHs), with masses in excess of a trillion solar masses have negative electric charge with an overall charge asymmetry of
about one in a billion billion.

\bigskip

\noindent
Structure formation in galaxies and clusters, including
the local void, is due only to gravitational forces. On the other hand, the structure formation regarding PEMBHs is due to electromagnetic forces, and its results regarding voids are expected to be quite different. In particular, what is the local void in terms of galaxies is nevertheless expected to contain PEMBHs whose electric charge can underly the origin of the Amaterasu cosmic ray, as follows.

\bigskip

\noindent
Consider a Primordial Extremely Massive Black Hole (PEMBH)
with mass $M_{PEMBH}=10^{12}M_{\odot}$ and negative electric
charge $q_{PEMBH}=-10^{32}$ Coulombs at a distance 1Mpc from the
Earth. Consider also an antiproton $\bar{p}$ at  rest. a candidate for
the Amaterasu primary, at a distance x metres for the Earth and aligned
between the PEMBH and the Earth. To be justified {\it a posteriori} we
assume that x metres $<<$ 1 Mpc.

\bigskip

\noindent
The Coulomb repulsion between PEMBH and $\bar{p}$ is given by
\begin{equation}
F = \frac{ k_e q_{PEMBH} q_{\bar{p}}}{r^2}
\end{equation}
where the electric force constant is $k_e = 9 \times 10^9 N.m^2/C^2$.
Using $1Mpc = 3 \times 10^{22}m$ and antiproton charge
$-1.6\times 10^{-19}$ Coulombs gives a repulsive electric force
which is approximately constant if x is sufficiently small
\begin{equation}
F= 1.6 \times 10^{-22} N
\end{equation}
in Newtons $N \equiv kg.m/s^2$. Approximating the antiproton mass
as $m(\bar{p}) = m_0 = 1.6 \times 10^{-19}$ kg the initial acceleration
\begin{equation}
a_i = a(\beta_i=0) = \frac{F}{m_0} = 1.0 \times 10^5 m/s^2
\end{equation} 

\bigskip

\noindent
The required BKZ final relativistic velocity $\beta_f = v_f/c$ is given by
\begin{equation}
\frac{E_f}{m_0} = \frac{1}{\sqrt{1-\beta_f^2}} = \frac{2.4\times10^{20} eV}{938\times10^6 eV} = 2.56 \times 10^{11}
\end{equation}
so that
\begin{equation}
\beta_f^2 = 1 - 1.52 \times 10^{-23}
\label{finalacc}
\end{equation}

\noindent
For the relativistic acceleration of $\beta = v/c$ from
$\beta_i =0$ to $\beta_f = \sqrt{1-1.52 \times 10^{-23}}$
we may use the integral
\begin{equation}
\int \frac{dx}{\sqrt{1-x^2}} = \sin^{-1} x
\label{integral}
\end{equation}

\bigskip

\noindent
We can integrate the motion from rest at time $t=t_i$ to reaching
energy $2.4 \times 10^{20}$ eV at time $t=t_f$ using the acceleration
\begin{equation}
\frac{d^2 s}{d t^2} =  c \frac{d \beta}{dt} = \frac{F}{m(\beta)} = \frac{F}{m_0} \sqrt{1-\beta^2} =
a_i \sqrt{1-\beta^2}
\label{acc}
\end{equation}
with the initial acceleration $a_i = 10^5 m/s^2$ (see Page 1) and $c=3\times 10^8m/s$.

\bigskip

\noindent
Study of Eq.(\ref{acc}) confirms its consistency since $\beta_i < 1$
provided that $m_0 >0$ and hence $a_i <   \infty$ so that the massive primary
particle never reaches the speed of light.  As indicated by Eq.(\ref{finalacc}), however,
the speed upon striking the Earth's upper atmosphere is only slightly
below this maximum possible speed.

\bigskip

\noindent
Using the integral in Eq.(\ref{integral}) now give the result
\begin{equation}
\beta_f = \sin \left[ \frac{t_f - t_i}{3000s} \right]
\label{time}
\end{equation}

\bigskip

\noindent
Since $\beta_f < 1$, we deduce from Eq.(\ref{time}) 
that $(t_f-t_i) < 3000s$ which implies that the initial 
at-rest antiproton must be less than $10^9$ km
from the Earth which is within the Solar System.
This is a remarkable result !

\bigskip

\noindent
We note that this requires exact alignment of the Amaterasu primary
with the PEMBH-to-Earth direction which is expected only very
rarely. Nevertheless, it does suggest that this singular Amaterasu cosmic ray
which hit the Earth's atmosphere on May 27, 2021 can shed light 
on the theory of the visible universe at the highest length scales.

\bigskip

\noindent
Although we have studied an antiproton primary, a proton primary
is also explicable within the EAU-model simply by placing the
proton behind the Earth and redefining alignment by requiring
that the detector is aligned {\it between} PEMBH and primary.
Since the antiproton would generate abnormal air showers, relative
to the proton case, we favour a proton primary in the event\cite{abbasi}.
Needless to say, the proton case involves the identical calculation
as for the antiproton case contained in our text.

\bigskip

\noindent
Structure formation for the PEMBHs is, as already discussed, quite
different than for  galaxies and clusters and the Local Void, This is
because its dynamics are governed by electromagnetism. Based
on only the one known example, we are unable to estimate
reliably the number of PEMBHs in the Local Void and hence
the number of future super-GKZ cosmic rays arising from the same
mechanism. For every future such ultra high energy cosmic ray it
should be checked whether its direction is consistent with that of the
Local Void.

\bigskip

\noindent
{\it Conclusion}.
Cosmic rays have historically had a major r\^{o}le in particle
physics, such as the original discoveries of the positron
and the pion. The Amaterasu cosmic ray is only one event but
it is an extraordinary one, as one of the three most energetic
cosmic rays ever recorded and the only one of those pointing
back to the Local Void where, according to the
$\Lambda CDM$ model, there is no obvious source.

\bigskip

\noindent
We have discussed a possible
explanation for the Amaterasu particle where the source is a PEMBH residing
in the Local Void which locally accelerates a proton primary.
In this scenario, a single exceptional cosmic ray has helped determine the correct choice
of theoretical cosmological model.

\bigskip

\noindent
{\bf Acknowledgement}

\noindent
We thank T.L. Curtright and T.W. Kephart for useful discussions, and the University of Salento, Lecce for an affiliation. We dedicate this work to the memory of our friend and colleague Tom Weiler, who
had an abiding interest in cosmic rays.


\bigskip


\begin{thebibliography}{999}

\bibitem{Greisen:1966jv}
  K.~Greisen,\\
{\it End to the cosmic ray spectrum?},\\
  Phys.\ Rev.\ Lett.\  {\bf 16}, 748 (1966).
  

\bibitem{Zatsepin:1966jv}
  G.~T.~Zatsepin and V.~A.~Kuzmin,\\
 {\it Upper Limit of the Spectrum of Cosmic Raya}.\\
  JETP Lett.\  {\bf 4}, 78 (1966).
 

\bibitem{abbasi}
R.~U.~Abbasi \textit{et al.} (The Telescope Array Collaboration),\\
{\it An Extremely Energetic Cosmic Ray Observed by a Surface Detector Drray,}\\
Science \textbf{382}, 903-907 (2023).\\
{\tt arXiv:2311.14231 [astro-ph.HE]}.

\bibitem{Anchordoqui:2018qom}
L.~A.~Anchordoqui,\\
{\it Ultra-High-Energy Cosmic Rays}.\\
Phys. Rept. \textbf{801}, 1-93 (2019).\\
{\tt arXiv:1807.09645 [astro-ph.HE]}.

\bibitem{FramptonPLB}
P.H. Frampton,\\
{\it Electromagnetic Accelerating Universe}.\\
Phys. Lett. {\bf B835,} 137480 (2022).\\
.{\tt arXiv:2210.10632[physics.gen-ph]}.

\bibitem{FramptonMPLA}
P.H. Frampton,\\
{\it  A Model of Dark Matter and Energy}.\\
Mod. Phys. Lett. {\bf A38,}  2350032 (2023).\\
{\tt arXiv:2301.10719[physics.gen-phys]}.

\bibitem{Einstein}
A. Einstein,\\
{\it Cosmological Considerations in the General Theory
of Relativity}.\\
Sitzungsber.Preuss.Akad.Wiss.Berlin. {\bf 1917,} 192 (1917).

\end{thebibliography}
\end{document}